# Chiral plasmons with twisted atomic bilayers

Xiao Lin[1,2], Zifei Liu[2], Tobias Stauber[3,4], Guillermo Gómez-Santos[5], Fei Gao[1,*], Hongsheng Chen[1], Baile Zhang[2,6,*], and Tony Low[7,*]

[1]*Interdisciplinary Center for Quantum Information, State Key Laboratory of Modern Optical Instrumentation, ZJU-Hangzhou Global Science and Technology Innovation Center, College of Information Science and Electronic Engineering, Zhejiang University, Hangzhou 310027, China.*
[2]*Division of Physics and Applied Physics, School of Physical and Mathematical Sciences, Nanyang Technological University, Singapore 637371, Singapore.*
[3]*Materials Science Factory, Instituto de Ciencia de Materiales de Madrid, CSIC, E-28049 Madrid, Spain.*
[4]*Institute for Theoretical Physics, University of Regensburg, D-93040 Regensburg, Germany.*
[5]*Departamento de Física de la Materia Condensada, Instituto Nicolás Cabrera and Condensed Matter Physics Center (IFIMAC), Universidad Autónoma de Madrid, E-28049 Madrid, Spain.*
[6]*Centre for Disruptive Photonic Technologies, NTU, Singapore 637371, Singapore.*
[7]*Department of Electrical and Computer Engineering, University of Minnesota, Minneapolis, Minnesota 55455, USA.*
[*]*e-mail: gaofeizju@zju.edu.cn (F. Gao); blzhang@ntu.edu.sg (B. Zhang); tlow@umn.edu (T. Low)*

**Van der Waals heterostructures of atomically thin layers with rotational misalignments, such as twisted bilayer graphene, feature interesting structural moiré superlattices. Due to the quantum coupling between the twisted atomic layers, light-matter interaction is inherently chiral; as such, they provide a promising platform for chiral plasmons in the extreme nanoscale. However, while the interlayer quantum coupling can be significant, its influence on chiral plasmons still remains elusive. Here we present the general solutions from full Maxwell equations of chiral plasmons in twisted atomic bilayers, with the consideration of interlayer quantum coupling. We find twisted atomic bilayers have a direct correspondence to the chiral metasurface, which simultaneously possesses *chiral* and *magnetic* surface conductivities, besides the common *electric* surface conductivity. In other words, the interlayer quantum coupling in twisted van der Waals heterostructures may facilitate the construction of various (e.g., bi-anisotropic) atomically-thin metasurfaces. Moreover, the chiral surface conductivity, determined by the interlayer quantum coupling, determines the existence of chiral plasmons and leads to a unique phase relationship (i.e., $\pm\pi/2$ phase difference) between their TE and TM wave components. Importantly, such a unique phase relationship for chiral plasmons can be exploited to construct the missing *longitudinal* spin of plasmons, besides the common *transverse* spin of plasmons.**

Chiral materials [1-10], which have left- and right-handed counterparts connected by the mirror symmetry, are ubiquitous in nature. Generally, natural chiral materials have weak chiral light-matter interactions. The emerging field of chiral plasmonics [11-16] exploits strong light-matter interactions for chiral optical responses, including giant circular dichroism and optical activity. As such, it holds great promise for chiral optical spectroscopy with enhanced sensitivity. Chiral surface plasmons are thus of fundamental importance in physics [1,8-10,17], biology [13,18], stereochemistry [4,13], and drug development [13,15] for discriminating chiral molecules of different handedness.

Van der Waals heterostructures with rotational misalignment [19-31] are a promising atomically-thin platform for chiral plasmonics. The finite twist and interlayer coupling between the atomic bilayers imply all mirror symmetry is broken, imparting the twisted van der Waals heterostructure with an inherent chirality [32-36]. Twisted van der Waals heterostructures also support the propagation of highly confined surface plasmons [37-43], thus providing an efficient way to strengthen chiral light-matter interactions in the extreme nanoscale [32-36]. Moreover, the active tunability of these atomic heterostructures is attractive for chiral plasmonics. First, the chemical potential of constituent two-dimensional (2D) materials [44] can be readily tuned, e.g., through electrostatic gating. As such, twisted van der Waals heterostructures can be an ideal platform for *active* chiral plasmonics [16]. Second, one can deterministically manipulate the interlayer twist angle, which dictates the electronic properties of van der Waals heterostructures [19-31], such as the recent discovery of superconducting and correlated insulating states in twisted bilayer graphene if the twist angle is close to the so-called "magic angle" [20-26].

Here we theoretically reveal the possibility to create the longitudinal spin of plasmons, in addition to their more well-known transverse spin, by exploiting the unique phase relationship (i.e., $\pm\pi/2$ phase difference) between the TE and TM wave components of chiral surface plasmons in twisted atomic bilayers. We highlight that while the longitudinal spin of photons and transverse spin of plasmons have been extensively studied [45-47], the transverse spin of photons was only recently reported [48], and the longitudinal spin of plasmons remains elusive. The longitudinal spin of plasmons might be of paramount importance to novel types of exotic spin-orbit interactions of light [46,49,50]. Moreover, we also show that stacked atomic bilayers with rotational misalignment can in general be effectively described by a chiral metasurface. The effective chiral metasurface simultaneously has chiral, magnetic and electric surface conductivities, where the chiral surface conductivity is the key parameter that determines the existence of chiral plasmons and their unique phase relationship mentioned above, and its value depends on the interlayer quantum coupling. From fundamental physical standpoint, the interlayer quantum coupling in twisted van der Waals heterostructures may facilitate the construction of various atomically-thin metasurfaces, ranging from isotropic, anisotropic, bi-isotropic to bi-anisotropic. Our work therefore bridges the gap between two seemingly disparate but exciting fields, namely metasurfaces and van der Waals heterostructures.

As a result of interlayer quantum coupling, twisted atomic bilayers possess an out-of-plane nonlocal response, namely $\sigma_{xy} \neq 0$ and $\bar{\bar{\sigma}}_1 \neq 0$, and from basic symmetry considerations such as time reversal symmetry and the Onsagar relation, their surface conductivity $\bar{\bar{\sigma}}_s$ can be expressed as a $4 \times 4$ matrix [34,35], namely

$$\bar{\bar{\sigma}}_s = \begin{bmatrix} \bar{\bar{\sigma}}_0 & \bar{\bar{\sigma}}_1 - i\sigma_{xy}\bar{\bar{\tau}}_y \\ \bar{\bar{\sigma}}_1 + i\sigma_{xy}\bar{\bar{\tau}}_y & \bar{\bar{\sigma}}_0 \end{bmatrix} \qquad (1)$$

where $\bar{\bar{\tau}}_y$ is the $2 \times 2$ Pauli matrix, respectively; $\bar{\bar{\sigma}}_0$, $\bar{\bar{\sigma}}_1$ and $\sigma_{xy}$ are the in-plane conductivity for each atomic layer, the covalent drag conductivity, and the Hall conductivity, respectively, which can be obtained, for example, via the continuum or tight-binding models [32-36]. Here, we restrict ourselves to the case where the atomic layer is isotropic; that is, $\bar{\bar{\sigma}}_0 = \sigma_0 \bar{\bar{1}}$ and $\bar{\bar{\sigma}}_1 = \sigma_1 \bar{\bar{1}}$ are diagonal, where $\bar{\bar{1}}$ is the unitary matrix. These conductivities, of course, depend on the twist angle $\theta$ [34,35]. In this work, we disregard any in-plane nonlocality. This is reasonable for cases where the wavelength of light or surface plasmons is much larger than the periodicity of Moire patterns in the twisted bilayers. In order to consider the out-of-plane nonlocal response of twisted atomic bilayers [Fig. 1(a)], they should be separated by a nonzero distance $a$ [34,35]. The interlayer quantum coupling or the out-of-plane nonlocal response is sensitive to $a$. For twisted bilayer graphene, $a \approx a_0$, where $a_0 = 0.35$ nm.

We now proceed to the discussion of electromagnetic boundary conditions for twisted atomic bilayers. According to the electromagnetic wave theory [51], the boundary conditions at the two atomic layers or at $z = \pm a/2$ [Fig. 1(a)] can be written as

$$\hat{n} \times \left( \bar{E}_{-}^{(1 \text{ or } 2)} - \bar{E}_{+}^{(1 \text{ or } 2)} \right) = 0 \tag{2}$$

$$\hat{n} \times \left( \bar{H}_{-}^{(1 \text{ or } 2)} - \bar{H}_{+}^{(1 \text{ or } 2)} \right) = \bar{J}_{s}^{(1 \text{ or } 2)} = \bar{\bar{\sigma}}_{s} \begin{bmatrix} \bar{E}_{-}^{(1)} \\ \bar{E}_{+}^{(2)} \end{bmatrix} \tag{3}$$

$\bar{E}_{\pm}^{(1 \text{ or } 2)}$ and $\bar{H}_{\pm}^{(1 \text{ or } 2)}$ stand for the components of electric and magnetic fields parallel to the interface [Fig. 1(a)], respectively. Their superscript 1 or 2 indicates whether the field is at the 1st or 2nd atomic layer, and their subscript − or + indicates whether the field is in the region above or below the corresponding interface. $\bar{J}_{s}^{(1 \text{ or } 2)}$ stands for the surface current. We highlight that equations (2-3) contain 8 coupled equations, since the off-diagonal terms of $\bar{\bar{\sigma}}_{s}$ in equation (3), which characterize the interlayer electromagnetic coupling, couples the fields of the two layers.

It is then desirable to simplify the boundary conditions in equations (2-3). After some simplification, we find that the boundary conditions can be re-organized into a more intuitive way as follows [Fig. 1(b)]

$$\hat{n} \times \left( \bar{E}_{-}^{(1)} - \bar{E}_{+}^{(2)} \right) = -\bar{\bar{\sigma}}_{m}(\bar{H}_{-}^{(1)} + \bar{H}_{+}^{(2)}) + \bar{\bar{\sigma}}_{\chi}(\bar{E}_{-}^{(1)} + \bar{E}_{+}^{(2)}) \tag{4}$$

$$\hat{n} \times \left( \bar{H}_{-}^{(1)} - \bar{H}_{+}^{(2)} \right) = +\bar{\bar{\sigma}}_{e}(\bar{E}_{-}^{(1)} + \bar{E}_{+}^{(2)}) + \bar{\bar{\sigma}}_{\chi}(\bar{H}_{-}^{(1)} + \bar{H}_{+}^{(2)}) \tag{5}$$

$$\bar{\bar{\sigma}}_{m} = \begin{bmatrix} \sigma_{m,x} & 0 \\ 0 & \sigma_{m,y} \end{bmatrix} = \begin{bmatrix} -\dfrac{f_{\text{TBG}}}{1-(\sigma_0-\sigma_1)\cdot f_{\text{TBG}}} & 0 \\ 0 & -\dfrac{f_{\text{TBG}}\cdot k_{z0}^2/k_0^2}{1-(\sigma_0-\sigma_1)\cdot f_{\text{TBG}}\cdot k_{z0}^2/k_0^2} \end{bmatrix} \tag{6}$$

$$\bar{\bar{\sigma}}_{\chi} = \begin{bmatrix} \sigma_{\chi,x} & 0 \\ 0 & \sigma_{\chi,y} \end{bmatrix} = -\sigma_{xy}\cdot \bar{\bar{\sigma}}_{m} \tag{7}$$

$$\bar{\bar{\sigma}}_{e} = \begin{bmatrix} \sigma_{e,x} & 0 \\ 0 & \sigma_{e,y} \end{bmatrix} = \begin{bmatrix} \sigma_0 + \sigma_1 - \sigma_{xy}^2 \sigma_{m,x} - \dfrac{\varepsilon_0}{\mu_0} f_{\text{TBG}} & 0 \\ 0 & \sigma_0 + \sigma_1 - \sigma_{xy}^2 \sigma_{m,y} - \dfrac{\varepsilon_0}{\mu_0} f_{\text{TBG}}\cdot \dfrac{k_{z0}^2}{k_0^2} \end{bmatrix} \tag{8}$$

In these equations, the coefficient $f_{\text{TBG}} = \dfrac{\omega \mu_0}{k_{z0}} \cdot \dfrac{e^{ik_{z0}a/2}-e^{-ik_{z0}a/2}}{e^{ik_{z0}a/2}+e^{-ik_{z0}a/2}}$ originates from the electromagnetic response of the gap between the two atomic layers (i.e., region 3 in Fig. 1, which can be readily treated as free space), where $k_{z0} = \sqrt{k_0^2 - q^2}$, $k_0^2 = \omega^2 \mu_0 \varepsilon_0$, $\mu_0$ and $\varepsilon_0$ are the permeability and permittivity of free space, respectively. Here we let the in-plane wavevector $\bar{q}$ to be parallel to the $+\hat{x}$ direction, i.e., $\bar{q} = \hat{x} k_x$. There is a factor difference of $\frac{k_{z0}^2}{k_0^2}$ between the $x$ and $y$ components of $\bar{\bar{\sigma}}_e$, $\bar{\bar{\sigma}}_m$ and $\bar{\bar{\sigma}}_\chi$ in equations (6-8). Such a difference comes from the fact that we have $\hat{z} \frac{\partial}{\partial z} \times \hat{y} E_y = i\omega\mu_0 \hat{x} H_x$ for TE (i.e., $s$-polarized) waves but $\hat{z} \frac{\partial}{\partial z} \times \hat{x} E_x = \frac{k_{z0}^2}{k_0^2} \cdot i\omega\mu_0 \hat{y} H_y$ for TM ($p$-polarized) waves within the gap between the two atomic layers in Fig. 1.

Now the simplified effective boundary conditions in equations (4-5) only contain 4 equations. To be specific, the effective boundary conditions [Fig. 1(b)] only relate to the fields outside the twisted bilayer (i.e., $|z| \geq a/2$), namely $\bar{E}_-^{(1)}$, $\bar{E}_+^{(2)}$, $\bar{H}_-^{(1)}$ and $\bar{H}_+^{(2)}$, while the fields inside the interlayer gap ($|z| \leq a/2$), namely $\bar{E}_+^{(1)}$, $\bar{E}_-^{(2)}$, $\bar{H}_+^{(1)}$ and $\bar{H}_-^{(2)}$, are eliminated. It is interesting to note that the effective boundary conditions satisfy the duality principle [51]. That is, equations (4-5) are duals of each other. If we make the following replacements, namely $\bar{E} \to \bar{H}$, $\bar{H} \to -\bar{E}$, $\bar{\bar{\sigma}}_m \to \bar{\bar{\sigma}}_e$, $\bar{\bar{\sigma}}_e \to \bar{\bar{\sigma}}_m$ and $\bar{\bar{\sigma}}_\chi \to \bar{\bar{\sigma}}_\chi$, then equation (4) becomes equation (5), and equation (5) becomes equation (4).

The obtained expression for $\bar{\bar{\sigma}}_e$, $\bar{\bar{\sigma}}_m$ and $\bar{\bar{\sigma}}_\chi$ in equations (4-8) holds for arbitrary propagation direction, since equations (4-8) is obtained by defining the $+\hat{x}$ direction to be the direction of in-plane wavevector $\bar{q}$. Meanwhile, since we are considering atomic layers which are isotropic, $\bar{\bar{\sigma}}_e$, $\bar{\bar{\sigma}}_m$ and $\bar{\bar{\sigma}}_\chi$ are all $2 \times 2$ diagonal matrices. Moreover, due to the different electromagnetic responses of the twisted bilayer to different polarizations of light (i.e., TE and TM waves), we generally have $\sigma_{e,x} \neq \sigma_{e,y}$, $\sigma_{m,x} \neq \sigma_{m,y}$ and $\sigma_{\chi,x} \neq \sigma_{\chi,y}$. Such a situation also happens to an *isotropic* dielectric slab. To be specific, if there are no twisted atomic bilayers in Fig. 1, namely $\sigma_0 = \sigma_1 = \sigma_{xy} = 0$ in equations (6-8), the thin isotropic dielectric slab (i.e., free space or region 3 in Fig. 1) has $\bar{\bar{\sigma}}_{m,\text{air}} = \begin{bmatrix} \sigma_{m,x,\text{air}} & 0 \\ 0 & \sigma_{m,y,\text{air}} \end{bmatrix}$, $\bar{\bar{\sigma}}_{e,\text{air}} = \frac{\varepsilon_0}{\mu_0} \bar{\bar{\sigma}}_{m,\text{air}}$, and $\bar{\bar{\sigma}}_{\chi,\text{air}} = 0$, where $\sigma_{m,x,\text{air}} = -f_{\text{TBG}}$ and $\sigma_{m,y} = \sigma_{m,x,\text{air}} \cdot k_{z0}^2/k_0^2$. Accordingly, the effective boundary conditions in equations (4-5) demonstrate that twisted atomic bilayers can be treated as a 2D bi-isotropic chiral metasurface [Fig. 1(b)]. In contrast, we highlight that in the original boundary conditions in equations (2-3), twisted atomic bilayers with the out-of-plane nonlocal response indeed shall be treated as a 3D material, due to its nonzero interlayer distance $a$. Without loss of generality, we denote $\bar{\bar{\sigma}}_e$, $\bar{\bar{\sigma}}_m$ and $\bar{\bar{\sigma}}_\chi$ in equations (6-8) as the effective electric, magnetic and chiral surface conductivities, respectively.

The dimensionless parameter $\bar{\bar{\sigma}}_\chi$ characterizes the magnetoelectric effect of the effective 2D chiral metasurfaces, instead of 3D bulk materials. Equation (7) shows that $\bar{\bar{\sigma}}_\chi$ depends on the Hall conductivity $\sigma_{xy}$. This way, the effective boundary conditions provide us a direct mapping of the interlayer quantum coupling of twisted atomic bilayers to the general constitutive parameters in Maxwell equations. If the interlayer distance for the two parallel atomic layers is relatively large [e.g., $a \gg a_0$ in Fig. 1(c, d)], the interlayer quantum coupling would diminish, which leads to $\sigma_{xy} = 0$ and thus $\bar{\bar{\sigma}}_\chi = 0$ from equation (7), even though the two atomic layers are rotationally misaligned. As such, our derived effective boundary conditions with $\bar{\bar{\sigma}}_\chi \neq 0$ are also an important extension of conventional boundary conditions with $\bar{\bar{\sigma}}_\chi = 0$ [51] for the electromagnetic wave theory itself.

We note that the forms of $\bar{\bar{\sigma}}_e$, $\bar{\bar{\sigma}}_m$ and $\bar{\bar{\sigma}}_\chi$, whose explicit expressions are given by equations (4-5), were derived for the case of non-magnetic and isotropic atomic bilayers. More general forms of $\bar{\bar{\sigma}}_e$, $\bar{\bar{\sigma}}_m$ and $\bar{\bar{\sigma}}_\chi$ can be realized by stacking other 2D atomic layers (e.g., in-plane anisotropic 2D materials, magnetic 2D materials) [28-30,44,52] or in conjunction with standard metasurface approaches such as nano-patterning [53,54]. With this in mind, below we explore surface plasmons governed by the effective boundary conditions in equations (4-5), where $\bar{\bar{\sigma}}_e$, $\bar{\bar{\sigma}}_m$ and $\bar{\bar{\sigma}}_\chi$ are taken to be arbitrary values. For conceptual demonstration, we assume the bi-isotropic metasurface to be located inside a symmetric dielectric environment. That is, regions 1 and 2 [Fig. 1(a, b)], which are the regions above and below the metasurface, respectively, have a relative permittivity of $\varepsilon_{1r} = \varepsilon_{2r} = \varepsilon_r > 0$.

For arbitrary values of $\bar{\bar{\sigma}}_\chi$, $\bar{\bar{\sigma}}_e$ and $\bar{\bar{\sigma}}_m$, the dispersions of surface plasmons is governed by

$$\frac{k_z^2}{\omega^2 \mu_0 \varepsilon_0 \varepsilon_r} \cdot \sigma_{\chi,x}^2 + \left(1 + \frac{k_z}{\varepsilon_r} \cdot \frac{\sigma_{e,x}}{\omega \varepsilon_0}\right) \cdot \left(1 + k_z \cdot \frac{\sigma_{m,x}}{\omega \mu_0}\right) = 0 \qquad (9)$$

$$\sigma_{\chi,y}^2 + \left(\frac{k_z}{\omega\mu_0} + \sigma_{e,y}\right) \cdot \left(\frac{k_z}{\omega\varepsilon_0\varepsilon_r} + \sigma_{m,y}\right) = 0 \tag{10}$$

where $k_z = \sqrt{\omega^2\mu_0\varepsilon_0\varepsilon_r - q^2}$ is the component of wavevector perpendicular to the interface. If $\bar{\bar{\sigma}}_\chi = 0$, equations (9-10) represent the dispersion for linearly-polarized surface plasmons. Moreover, if $Im(\sigma_{e,x}) > 0$ [44] or $Im(\sigma_{m,y}) < 0$ in equations (9-10), TM surface plasmons exist. Similarly, if $Im(\sigma_{e,y}) < 0$ [55] or $Im(\sigma_{m,x}) > 0$, TE surface plasmons can appear.

If $\bar{\bar{\sigma}}_\chi \neq 0$, equations (9-10) stand for the dispersion for chiral plasmons. Due to the magnetoelectric effect, these chiral plasmons are intrinsically hybrid TE-TM waves regardless of the values of $\bar{\bar{\sigma}}_e$ and $\bar{\bar{\sigma}}_m$. As such, chiral plasmons with $\bar{\bar{\sigma}}_\chi \neq 0$ are distinctly different from the linearly-polarized surface plasmons with $\bar{\bar{\sigma}}_\chi = 0$. Particularly, if $\bar{\bar{\sigma}}_e = 0$ and $\bar{\bar{\sigma}}_m = 0$ in equations (9-10), the dispersion of the two eigenmodes of chiral plasmons is reduced to

$$\frac{k_z^2}{\omega^2\mu_0\varepsilon_0\varepsilon_r} \cdot \sigma_{\chi,x}^2 + 1 = 0 \tag{11}$$

$$\frac{k_z^2}{\omega^2\mu_0\varepsilon_0\varepsilon_r} + \sigma_{\chi,y}^2 = 0 \tag{12}$$

Figure 2 shows the basic features of these chiral plasmons governed by equations (11-12), including their dispersion (namely the relation between $q$ and $\sigma_{\chi,x}^2$ or $\sigma_{\chi,y}^2$) in Fig. 2(a) and their field distribution in Fig. 2(b, c). Figure 2(a) shows that we have $q/(\omega/c) \gg 1$ (namely large spatial confinement) for one chiral eigenmode in equation (11) if $\sigma_{\chi,x} \to 0$, while we have $q/(\omega/c) \to 1$ (small spatial confinement) for the other chiral eigenmode in equation (12) if $\sigma_{\chi,y} \to 0$. For the chiral eigenmodes in equation (11), Fig. 2(b) shows that their field distributions of both $H_z(\bar{r})$ and $E_z(\bar{r})$ are odd symmetric with respect to the interface (namely the plane of $z = 0$). Here $H_z(\bar{r})$ and $E_z(\bar{r})$ are the field components for the TE and TM waves, respectively. It is then straightforward to denote these modes in equation (11) as the chiral odd eigenmodes of surface plasmons. Similarly, the modes in equation (12) are denoted as the chiral even eigenmodes of surface plasmons, since their field distribution of both $H_z(\bar{r})$ and $E_z(\bar{r})$ are even symmetric with respect to the interface [Fig. 2(c)].

Moreover, we find in Fig. 2(b, c) and Fig. S2 that there is a uniqe phase relationship between the TE and TM field components for chiral plasmons. The phase difference between the TE and TM field components, namely $\text{Arg}(H_z(\bar{r})/E_z(\bar{r}))$, is dependent on the sign of $\sigma_{\chi,x}$ and $\sigma_{\chi,y}$. To be specific, we have $\text{Arg}\left(\frac{H_z(\bar{r})}{E_z(\bar{r})}\right) = -\frac{\pi}{2} \cdot \text{sgn}(\sigma_{\chi,y})$ for chiral even eigenmodes but $\text{Arg}\left(\frac{H_z(\bar{r})}{E_z(\bar{r})}\right) = +\frac{\pi}{2} \cdot \text{sgn}(\sigma_{\chi,x})$ for chiral odd eigenmodes.

Such unique phase relationships in chiral surface plasmons can be adopted to emulate the *longitudinal* spin of plasmons in Fig. 3(a, b), in addition to the common *transverse* spin of plasmons in Fig. 3(c) [45-47]. For the chiral surface plasmons in Fig. 2 in the $z < 0$ region, they have the electric field $\bar{E}_{1,\sigma_{\chi,y}} \propto (\hat{x}\frac{k_{z0}}{k_0} + \hat{y} \cdot (-i) \cdot \text{sgn}(\sigma_{\chi,y}) + \hat{z}\frac{k_x}{k_0})e^{ik_x x - ik_{z0} z}$ for the eigenmodes related to $\sigma_{\chi,y}$, where $k_{z0} = i\text{Im}(k_{z0})$. Similarly, they have the electric field $\bar{E}_{1,\sigma_{\chi,x}} \propto \left(\hat{x}\frac{k_{z0}}{k_0} + \hat{y} \cdot (+i) \cdot \text{sgn}(\sigma_{\chi,x}) + \hat{z}\frac{k_x}{k_0}\right)e^{ik_x x - ik_{z0} z}$ for the eigenmode related to $\sigma_{\chi,x}$. Then the electric-field ellipticity $\varphi_E = \text{Im}(\bar{E}^* \times \bar{E})$ (or spin [56,57]) both in the $\hat{x}$ and $\hat{y}$ directions would not vanish for chiral surface plasmons, and its gradient would locally produce the spin energy flow [47]. In other words, the cycloidal rotation of electric fields in the $y$-$z$ plane would generate the longitudinal spin $\bar{S}_x$ (parallel to $\bar{q}$), while the cycloidal rotation of electric fields in the $x$-$z$ plane would generate the transverse spin $\bar{S}_y$ (perpendicular to $\bar{q}$), as shown in Fig. 3(a, b). Moreover, if $\left|\frac{k_{z0}}{k_0}\right| \ll 1$ (e.g., chiral surface plasmons with $|\sigma_{\chi,y}| \ll 1$ in Fig. 2), we would have the cycloidal rotation of electric fields mainly in the $y$-$z$ plane, inducing the almost pure longitudinal spin of plasmons. In contrast, if $\left|\frac{k_{z0}}{k_0}\right| \gg 1$ (e.g., chiral surface plasmons with $|\sigma_{\chi,x}| \ll 1$ in Fig. 2), the cycloidal rotation of electric fields happens mainly in the $x$-$z$ plane, giving rise to the almost pure transverse spin of plasmons; see also the pure transverse spin for linearly polarized plasmons in Fig. 3(c).

Regarding the twisted bilayer whose constituent atomic layers are isotropic and non-magnetic (e.g., twisted bilayer graphene), we generally have $\bar{\bar{\sigma}}_\chi \neq 0$, $\bar{\bar{\sigma}}_e \neq 0$ and $\bar{\bar{\sigma}}_m \neq 0$ in equations (4-8). Then, their supported chiral plasmons have the dispersion governed by equations (9-10). As a typical example, we shows in Fig. S1 the dispersion of chiral plasmons for the case when the twisted atomic bilayers have $\sigma_0 = \sigma_1 = 0$ and $\sigma_{xy} \neq 0$. Under these specific conditions and if $\varepsilon_r = 1$, equations (9-10) can be further reduced into the following simple form,

$$[t^{+1/2} + t^{-1/2}]^2 - \frac{\mu_0}{\varepsilon_0} \cdot \sigma_{xy}^2 = 0 \tag{13}$$

where $t = \frac{-k_{z0}}{\omega \mu_0} \cdot f_{\text{TBG}}$. Since $|t^{+1/2} + t^{-1/2}| \geq 2$, there is a critical existence condition for these chiral plasmons governed by equation (13), namely $|\sigma_{xy}^2| \geq 4\varepsilon_0/\mu_0$. The in-plane wavevector of these chiral plasmons would go to infinity (i.e., extreme spatial confinement) if $|\sigma_{xy}|$ approaches to $\sqrt{4\varepsilon_0/\mu_0}$, and it would decrease to $\omega/c$ (poor spatial confinement) if $|\sigma_{xy}| \gg \sqrt{4\varepsilon_0/\mu_0}$, as shown in Fig. S1. Moreover, for twisted atomic bilayers such as twisted bilayer graphene, due to the flexible tunability of their effective conductivities via changing the chemical potential and twist angle, it can provide us an ideal platform to explore some other exotic features of chiral plasmons, which are worthy of further in-depth studies in the future.

In summary, we theoretically investigate basic features of chiral plasmons that can emerge in van der Waals heterostructures comprised of twisted atomic bilayers. As a unique emerging feature of chiral plasmons, we find that they possess the longitudinal spin of plasmons, besides to the common transverse spin of plasmons. The revealed longitudinal spin of plasmons may enable many other exotic spin-orbit interactions of light [45-50]. We have also directly mapped twisted atomic bilayers to the chiral metasurface, which simultaneously possesses the chiral, magnetic and electric surface conductivities. Correspondingly, the derived effective boundary conditions for twisted atomic bilayers can further deepen our understanding of the influence of interlayer quantum coupling on the chiral optical and plasmonic response. Meanwhile, the effective boundary conditions can enable a more intuitive and simplified way to treat light scattering problems in the new class of twisted atomic bilayers. Our work further indicates that chiral light-matter interactions in the extreme nanoscale can be efficiently tailored via the interlayer quantum coupling or the interlayer twisting. Considering a large library of atomically thin 2D materials, twisted van der Waals heterostructures can provide a fundamental building block to engineer various atomically-thin and actively tunable metasurfaces with their electromagnetic response ranging from isotropic, anisotropic, bi-isotropic to bi-anisotropic [51]. As such, twisted van der Waals heterostructures can also offer a fertile frontier for exploring the emerging field of chiral plasmonics.

**Acknowledgements**
This work was sponsored by Nanyang Technological University for NAP Start-Up Grant, the Singapore Ministry of Education (Grant No. MOE2015-T2-1-070, MOE2016-T3-1-006, and Tier 1 RG174/16 (S)), Spain's MINECO under Grant No. FIS2017-82260-P, PGC2018-096955-B-C42, and CEX2018-000805-M as well as by the CSIC Research Platform on Quantum Technologies PTI-001. This work at Zhejiang University was sponsored by the National Natural Science Foundation of China (NNSFC) under Grants No. 61801426, No. 61625502, No.11961141010, and No. 61975176, the Top-Notch Young Talents Program of China, the Zhejiang Provincial Natural Science Foundation under Grants No. Z20F010018, and the Fundamental Research Funds for the Central Universities.

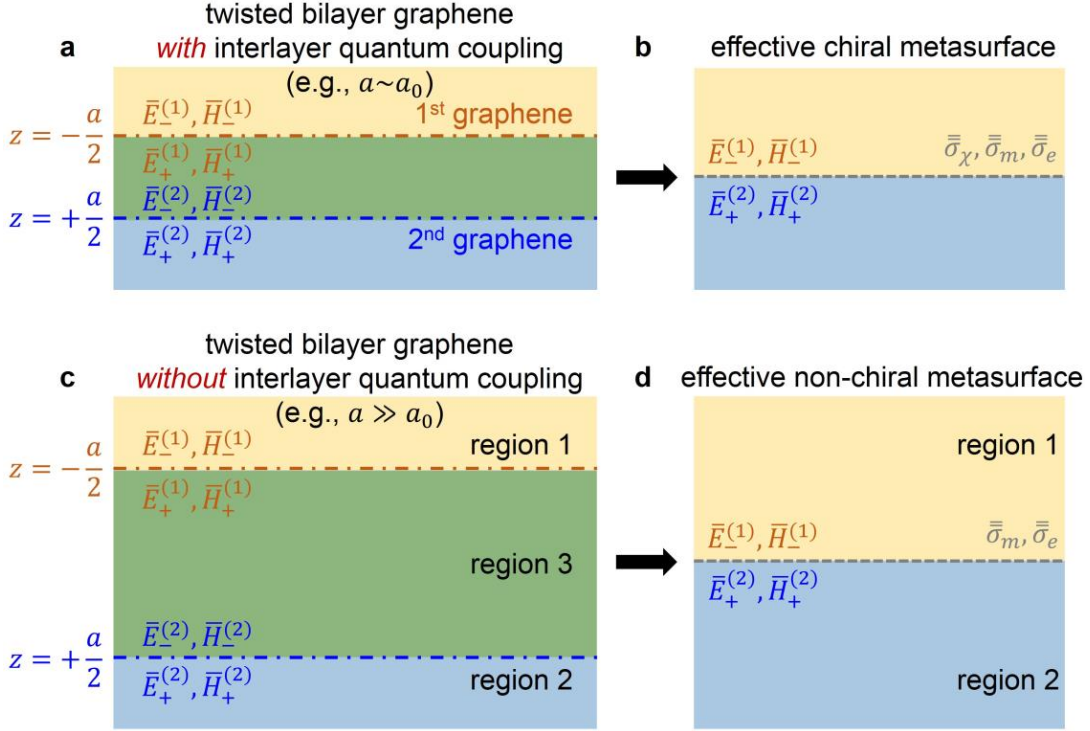

**Figure 1. Simplified effective electromagnetic boundary conditions for twisted atomic bilayers.** The twisted atomic bilayer is represented by twisted bilayer graphene in this plot. (**a, b**) For twisted bilayer graphene, its interlayer distance $a$ is comparable to $a_0 = 0.35$ nm, and it has nontrivial interlayer quantum coupling, which would induce the out-of-plane nonlocal response. (a) Original boundary conditions with the out-of-plane nonlocal description. (b) Effective boundary conditions, where the twisted atomic bilayer is equivalently modelled by a chiral metasurface. The chiral metasurface has an electric surface conductivity $\bar{\bar{\sigma}}_e$, a magnetic surface conductivity $\bar{\bar{\sigma}}_m$ and a chiral surface conductivity $\bar{\bar{\sigma}}_\chi$. (**c, d**) For hypothetical twisted atomic bilayers, if their interlayer distance is relatively large, namely $a \gg a_0$, the interlayer quantum coupling would disappear. Such hypothetical twisted atomic bilayers are equivalent to a non-chiral metasurface with $\bar{\bar{\sigma}}_\chi = 0$. Region 3 in (a, c) can be treated as free space.

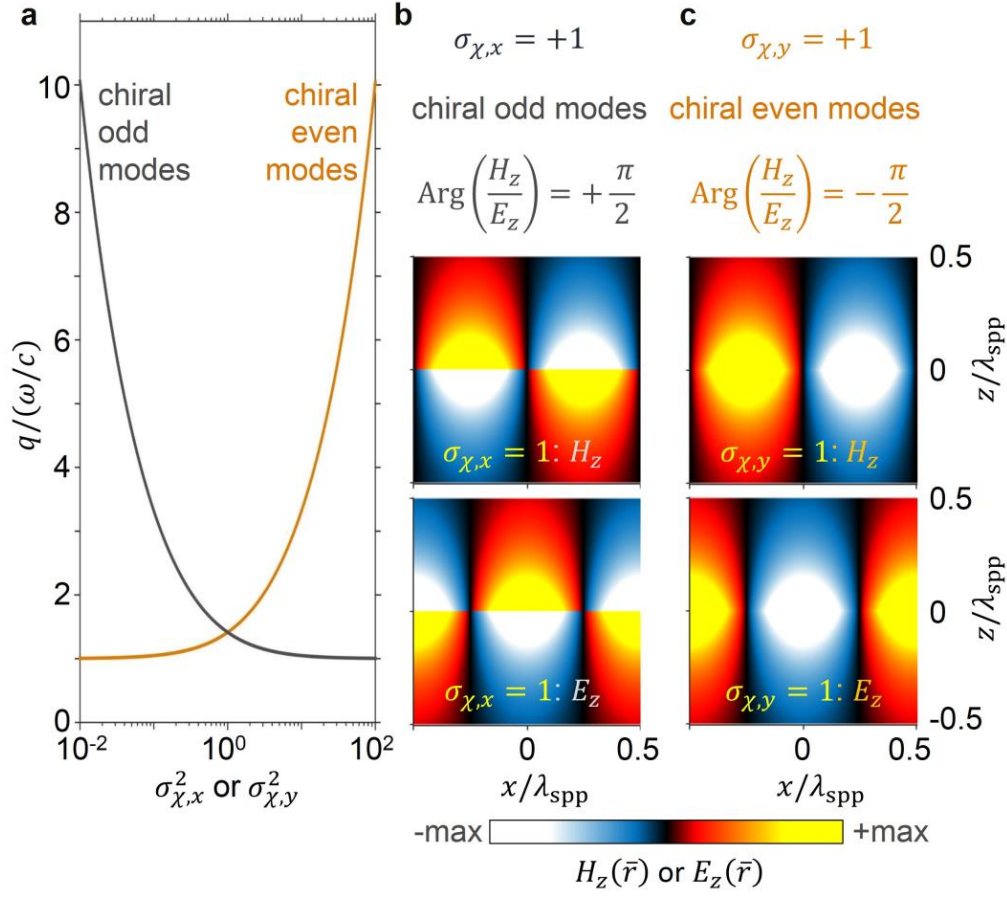

**Figure 2. Chiral surface plasmons supported by the chiral surface conductivity $\bar{\bar{\sigma}}_\chi$.** Here the chiral metasurface has $\bar{\bar{\sigma}}_\chi \neq 0$, $\bar{\bar{\sigma}}_m = 0$ and $\bar{\bar{\sigma}}_e = 0$, and it is located at $z = 0$ in free space. (**a**) In-plane wavevector $q$ of chiral surface plasmons as a function of $\sigma_{\chi,x}^2$ or $\sigma_{\chi,y}^2$. (**b, c**) Field distribution of chiral surface plasmons at (b) $\sigma_{\chi,x} = +1$ and (c) $\sigma_{\chi,y} = +1$. We have the phase difference between the TE and TM field components $\mathrm{Arg}\left(\frac{H_z(\bar{r})}{E_z(\bar{r})}\right) = -\frac{\pi}{2} \cdot \mathrm{sgn}(\sigma_{\chi,y})$ for chiral even eigenmodes but $\mathrm{Arg}\left(\frac{H_z(\bar{r})}{E_z(\bar{r})}\right) = +\frac{\pi}{2} \cdot \mathrm{sgn}(\sigma_{\chi,x})$ for chiral odd eigenmodes. Here the plasmonc wavelength $\lambda_{\mathrm{spp}}$ is equal to $2\pi/q$. The field distribution of chiral plasmons at $\sigma_{\chi,x} = -1$ and $\sigma_{\chi,y} = -1$ is shown in Fig. S2.

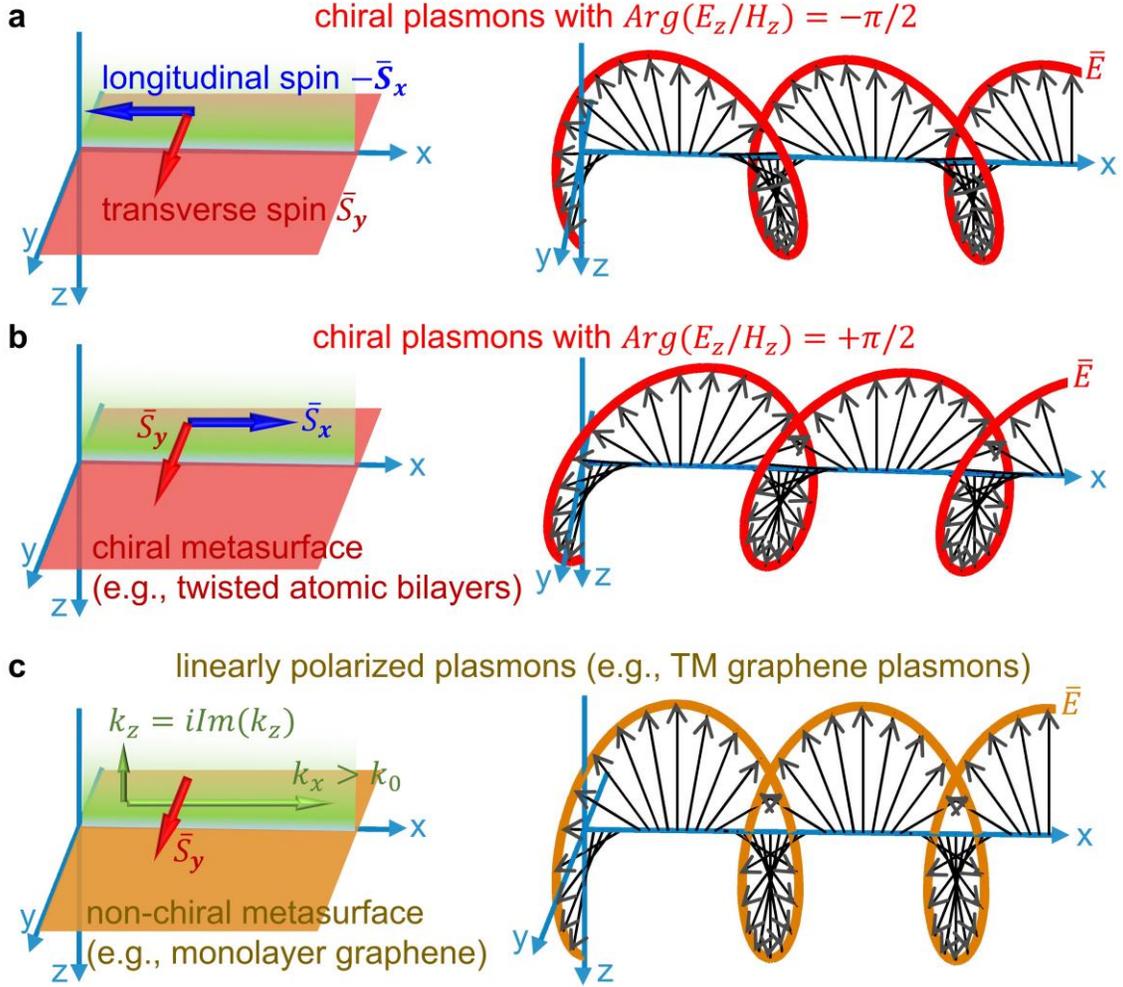

**Figure 3. Longitudinal spin in chiral surface plasmons.** The surface plasmons propagate along $+\hat{x}$ direction and decay exponentially away from the interface. The right-side panels show the instantaneous distribution of electric fields in the $z < 0$ region. (**a, b**) Chiral surface plasmons. The structural setup in (a) and (b) are the same as the chiral even eigenmdoe in Fig. 2(c) and the chiral odd eigenmode in Fig. 2(b), respectively. The cycloidal rotation of electric fields in the $y$-$z$ plane generates the longitudinal spin $\bar{S}_x$ (parallel to the in-plane wavevector $\bar{q} = \hat{x}k_x$ of chiral plasmons), while the cycloidal rotation of electric fields in the $x$-$z$ plane generates the transverse spin $\bar{S}_y$ (perpendicular to $\bar{q}$). (**c**) For comparison, we show the linearly polarized surface plasmons in (c), whose electric fields rotate only in the $x$-$z$ plane and possess only the transverse spin.